\documentclass[useAMS]{mn2e}
\usepackage{graphicx,graphics}
\usepackage{amsmath}

\title[Broadband spectroscopy of 4U~1700-37 with Suzaku]{Broadband spectroscopy of the eclipsing high mass X-ray binary 4U~1700-37 with Suzaku}

\author[G. K. Jaisawal \& S. Naik]{Gaurava K. Jaisawal$^1$\thanks{gaurava@prl.res.in} and Sachindra Naik$^1$\thanks{snaik@prl.res.in} \\
$^1$Astronomy and Astrophysics Division, Physical Research Laboratory, Navrangapura, Ahmedabad - 380009, Gujarat, India\\}

\begin{document}

\date{}

\maketitle

\begin{abstract}
We present the results obtained from broadband spectroscopy of the high mass 
X-ray binary 4U~1700-37 using data from a $Suzaku$ observation in 
2006 September 13-14 covering 0.29-0.72 orbital phase range. The light 
curves showed significant and rapid variation in source flux during entire 
observation. We did not find any signature of pulsations in the light curves. 
However, a quasi-periodic oscillation at $\sim$20 mHz was detected in the power 
density spectrum of the source. The 1-70 keV spectrum was fitted with various 
continuum models. However, we found that the partially absorbed high energy 
cutoff power-law and Negative and Positive power-law with Exponential cutoff 
(NPEX) models described the source spectrum well. Iron emission lines at 6.4 keV 
and 7.1 keV were detected in the source spectrum. An absorption like feature at 
$\sim$39 keV was detected in the residuals while fitting the data with NPEX model. 
Considering the feature as cyclotron absorption line, the surface magnetic field 
of the neutron star was estimated to be $\sim$3.4$\times$10$^{12}$ Gauss. To 
understand the cause of rapid variation in the source flux, time-resolved 
spectroscopy was carried out by dividing the observation into 20 narrow segments. 
The results obtained from the time-resolved spectroscopy are interpreted as the 
accretion of inhomogeneously distributed matter in the stellar wind of the supergiant 
companion star as the cause of observed flux variation in 4U~1700-37. A sharp increase 
in column density after $\sim$0.63 orbital phase indicates the presence of an accretion 
wake that blocks the continuum and produces the eclipse like low-flux segment.
\end{abstract}


\begin{figure*}
\centering
\includegraphics[height=5.5in, width=4.2in, angle=-90]{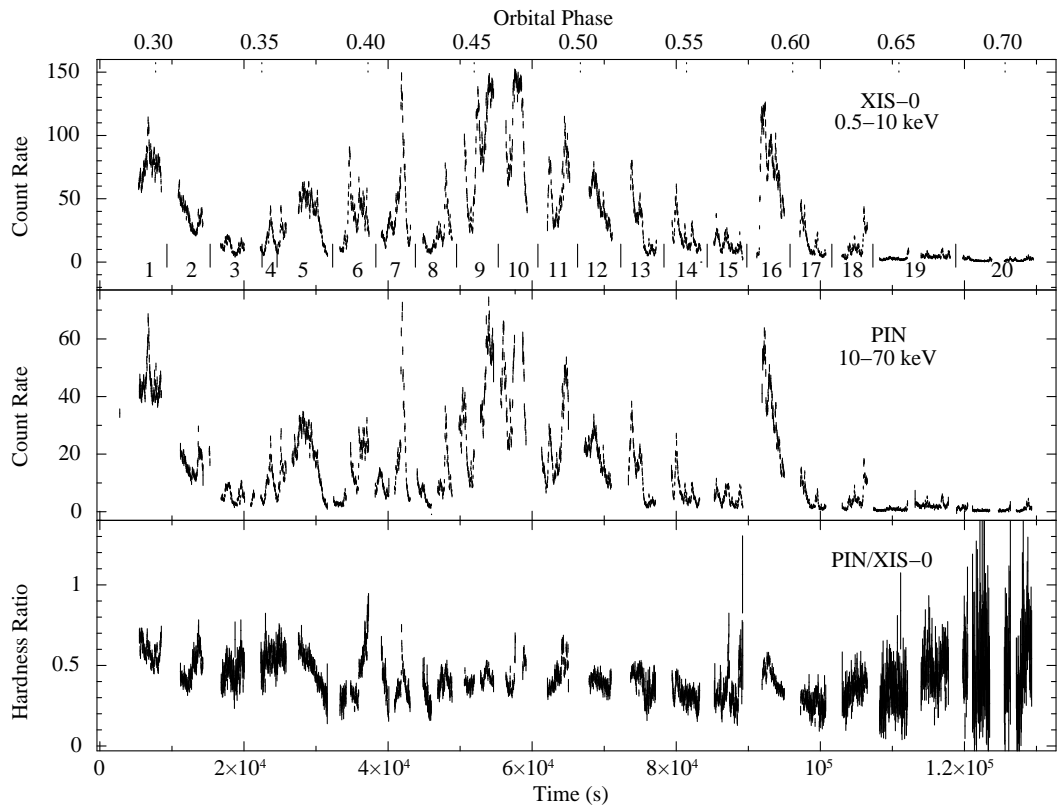}
\caption{Light curves (top and middle panels) and hardness ratio (bottom 
panel) obtained from the {\it Suzaku} observation of the high-mass X-ray 
binary 4U~1700-37. Data from XIS-0 and HXD/PIN detectors are plotted here. 
Flux variability by an order of $\sim$10-15 can be seen in top and middle 
panels of the figure. The quoted numbers at the top panel show the duration 
of segments used for time-resolved spectroscopy. The orbital phases covered 
during the {\it Suzaku} observation are marked at the top of the figure.}
\label{lc}
\end{figure*}

\section{Introduction}

4U~1700-37 was discovered by $Uhuru$ satellite in December 1970 (Jones et al. 1973). 
Extensive follow-up observations with $Uhuru$ revealed the system as an eclipsing 
binary with an orbital period of 3.412 days. One of the most luminous and hottest 
optical star among the known high mass X-ray binaries, a supergiant star (HD~153919) 
of O6.5 Iaf spectral type was identified as the optical companion (Hutchings et al. 
1973). Using $BATSE$ data, the orbital parameters of the binary system such as 
inclination $i$=66$^\circ$, eccentricity $e<$ 0.01, 48 $<$ a$_x$ sin~$i < $82 lt-sec 
and semi eclipse angle $\theta_E$=28$^\circ$.6 were derived (Rubin et al. 1996). Using
Monte Carlo simulation, the mass of compact object and mass and radius of the optical 
companion star were constrained at M$_x \sim$ 2.6 M$_\odot$, M$_o \sim$ 30 M$_\odot$ 
and R$_o \sim$ 18 R$_\odot$, respectively (Rubin et al. 1996). Using ultraviolet and
optical spectroscopic observations, Clark et al. (2002) evaluated physical parameters 
of the optical companion and used in Monte Carlo simulation to estimate the mass of the
X-ray source and optical companion to be M$_x\sim$2.44$\pm$0.27 M$_\odot$ and 
M$_o\sim$58$\pm$11 M$_\odot$. The distance of the binary system was estimated to be 
1.9 kpc (Ankay et al. 2001).

A tentative detection of pulsation at $\sim$67~s was reported from $Tenma$ 
observations of 4U~1700-37 (Murakami et al. 1984). However, later observations 
did not confirm the detection of spin period in the source. Although detection of 
X-ray pulsations are not yet confirmed, the spectrum of 4U~1700-37 has been 
well described by the standard continuum models of the accretion powered X-ray pulsars. 
Broad-band X-ray spectrum of 4U~1700-37 obtained from various observatories such as
$HEAO~1$, $EXOSAT$, $Ginga$, {\it Beppo}SAX had been described with a high energy cutoff 
power-law model (White et al. 1983; Haberl et al. 1989; Haberl \& Day 1992; Reynolds 
et al. 1999). A soft excess component was also detected in the spectrum during the 
eclipse and eclipse ingress observations of 4U~1700-37 (Haberl et al. 1989; Haberl 
\& Day 1992). 1991 April $Ginga$ observation of the source showed a clear difference in the 
temperature corresponding to the soft excess component before and after the eclipse 
(0.47 keV and 0.74 keV). The rise in soft excess temperature after the eclipse was 
explained on the basis of the bow shock formation in front of compact object (Haberl 
et al. 1994).

The nature of the compact object in 4U~1700-37 binary system is quite unclear . 
Due to observed hard X-ray spectrum and non-detection of pulsation, Brown et al. 
(1996) suggested the X-ray source as a low mass black hole candidate. However, 
the 2-200 keV {\it Beppo}SAX spectrum of 4U~1700-37 was found to be well fitted 
with a high energy cutoff power-law model representing the classical spectrum of 
an accretion powered X-ray pulsars (Reynolds et al. 1999). Apart from this, 
Reynolds et al. (1999) also reported the possible presence of cyclotron absorption 
feature at $\sim$37 keV in {\it Beppo}SAX spectrum. These results discarded the 
possibility of a black hole as a compact object in the 4U~1700-37 binary system. 
High resolution spectra from $Chandra$ and $XMM-Newton$ observatories were described 
by using two component absorption model as used in 1991 April $Ginga$ observation 
(Boroson et al. 2003; van der Meer et al. 2005). The recombination lines from H and 
He like species and fluorescence emission lines from neutral atoms were seen in the 
$Chandra$ observations during intermittent flare state of 4U~1700-37. The strength 
of the lines were found to be varying over the observation. The detection of triplet 
structure in Si and Mg indicated the disequilibrium of the photo-ionized plasma 
(Boroson et al. 2003). Many recombination lines as well as fluorescence emission 
lines were also detected in the spectrum of $XMM-Newton$ observation of 4U~1700-37 
during eclipse, eclipse egress and low flux segments of the binary (van der Meer et 
al. 2005). Presence of recombination lines from H and He atoms in eclipse phase 
suggested an extended ionization region around the source. As in the case of accretion 
powered X-ray pulsars, several mHz QPOs were also detected in the power density spectra 
of 4U~1700-37, obtained from the $Chadra$ observation (Boroson et al. 2003).

4U~1700-37 was observed by $Suzaku$ in 2006 September during out of eclipse phase 
of the binary. We report the time-averaged and time-resolved broadband spectroscopy 
of 4U~1700-37 to understand the nature of the continuum emission and its orbital 
dependency, flaring activities, emission lines and cyclotron features in the spectrum.

\section{Observation and Analysis}
  
$Suzaku$ is the fifth Japanese X-ray satellite which was launched in 
2005 July by Japan Aerospace Exploration Agency (Mitsuda et al. 2007). 
It covers a broad energy range (from 0.2~keV to 600~keV) in the X-ray band 
with the help of two sets of detectors, X-ray Imaging Spectrometers (XIS; 
Koyama et al. 2007) and Hard X-ray Detectors (HXD; Takahashi et al. 2007). 
XISs are imaging CCD cameras that work in 0.2-12 keV range. Three CCD cameras 
(XIS-0, XIS-2 and XIS-3) are front-illuminated whereas the other one (XIS-1) 
is back-illuminated. The effective areas of front-illuminated and 
back-illuminated XISs are 340~cm$^2$ and 390~cm$^2$ at 1.5~keV, 
respectively. Field of view of XIS is $18'\times18'$ in full window 
mode. HXD is a non-imaging detector consisting of two instruments 
such as silicon PIN diodes (HXD/PIN) and GSO crystal scintillators 
(HXD/GSO) working in 10-70~keV and 40-600~keV ranges, respectively. 
Effective area of the HXD/PIN is 145 cm$^2$ at 15 keV whereas for 
GSO, it is 315 cm$^2$ at 100 keV. Field of view of HXD/PIN is 
$34'\times34'$ and is similar for HXD/GSO up to 100 keV.

4U~1700-37 was observed with $Suzaku$ in 2006 September 13-14. The 
observation was carried out during out of eclipse phase of the binary
covering 0.29-0.72 orbital phase range (considering mid-eclipse time
as phase zero; Rubin et al. 1996). The observation was performed in 
``XIS nominal'' position with an effective exposure of $\sim$81.5~ks 
and $\sim$82.1~ks for XIS and HXD, respectively. XIS detectors were 
operated in the ``burst'' clock mode with ``1/4 window'' option 
providing 1~s time resolution during the observation. We used publicly 
available data (version 2.0.6.13) of the {\it Suzaku} observation
in the present work. HEASoft software package (version 6.12) and 
calibration database (CALDB) released on 2012 February 10 (for XIS) 
and 2011 September 13 (for HXD) were used for the data analysis. 

Unfiltered event files were processed by using `aepipeline' package of 
FTOOLS along with standard screening criteria to create cleaned XIS and 
PIN event files. These reprocessed clean event files were used in further 
analysis. ``XSELECT'' package of FTOOLS was used to extract light curves 
and spectra from the reprocessed XIS and HXD/PIN event data. Barycentric 
correction was applied on the reprocessed clean event files by using the 
task `aebarycen'. Attitude correction was applied to the reprocessed XIS 
event data files by using S-lang script 
\textit{aeattcor.sl}\footnote{http://space.mit.edu/ASC/software/suzaku/aeattcor.sl}. 
The reprocessed XIS event data were checked for the possible presence of photon 
pile-up. Photon pile-up in XIS event data was estimated by using S-lang script 
\textit{pile\_estimate.sl}\footnote{http://space.mit.edu/ASC/software/suzaku/pile\_estimate.sl} 
and found to be 10\%, 11\%, 11\% and 11\% at the center of the image obtained from 
XIS-0, XIS-1, XIS-2 and XIS-3 event data, respectively. An annulus region with inner 
and outer radii of $30''$  and $180''$ from the source position was selected to reduce 
the photon pile-up to $\leq$4\%. Source light curves and spectra were extracted from 
the reprocessed XIS event data by selecting above annulus region around the central 
source. Background light curves and spectra were accumulated by selecting circular 
regions away from the source. The response and effective area files for all the XIS 
detectors were generated by using the task `xisrmfgen' and `xissimarfgen' of FTOOLS. 
Source light curve and spectra were created from reprocessed HXD/PIN event file by 
using ``XSELECT''. However, the HXD/PIN background light curves and spectrum were 
accumulated in a similar manner from `tuned' non X-ray background 
(NXB\footnote{http://heasarc.nasa.gov/docs/suzaku/analysis/pinbgd.html}) event file. 
A correction for cosmic X-ray background 
(CXB\footnote{http://heasarc.nasa.gov/docs/suzaku/analysis/pin\_cxb.html}) was incorporated 
in the PIN spectra as suggested by the instrument team. Epoch 2 response file (20080129) 
for HXD/PIN was used in the spectral analysis. Data from all four XISs (XIS-0, XIS-1, XIS-2 
and XIS-3) and HXD/PIN were used in the present study.

\begin{figure}
\centering
\includegraphics[height=3.2in, width=2.5in, angle=-90]{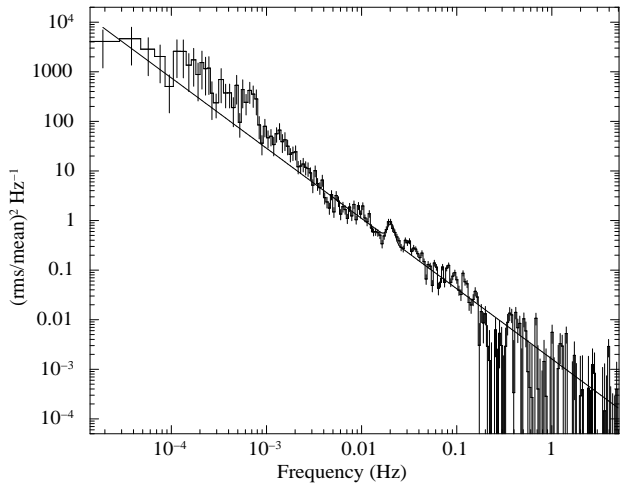}
\caption{Power density spectrum (PDS) of 4U~1700-37 obtained
from HXD/PIN light curve in 10-70 keV energy range. Absence of
pulsations in the range of 10$^{-5}$ Hz to 2 Hz range can be
seen in the figure. A QPO at $\sim$20 mHz is detected in the PDS of
the X-ray source. The solid line in the figure represents the fitted 
model comprising of a power-law continuum and a Gaussian function at
the QPO frequency.}
\label{qpo}
\end{figure}

\section{Results}

Source and background light curves in soft (XIS - 1 s time resolution) and 
hard X-ray (HXD/PIN - 0.1 s time resolution) energy ranges were extracted as 
described above. Background subtracted light curves in 0.5-10 keV and 10-70 keV 
ranges are shown in top and middle panels of Figure~\ref{lc}. From the figure,
significant and rapid flux variability by a factor of $\sim$10-15 can be seen
in soft and hard X-ray bands. The presence of flaring episodes along with stable 
low flux segments can also be clearly seen in XIS and PIN light curves. Hardness 
ratio (ratio between the light curves obtained from HXD/PIN and XIS-0 event data) 
plot (bottom panel of Figure~\ref{lc}) was generated to check the spectral state 
of the source during the flaring episodes as well as low flux segments in the light 
curve. However, apart from marginal hardening during the extended low flux segment 
towards the end of the observation, any significant change in the value of hardness 
ratio (spectral state) was absent. 

To investigate the presence of any periodicity (pulsation) in the X-ray source,
power density spectrum (PDS) was generated by using HXD/PIN light curve with 
0.1 s time resolution and shown in Figure~\ref{qpo}. Absence of any clear and sharp 
peaks in the PDS in 0.5~s to 10$^5$~s range suggested the non-detection of pulsation
in above time range. To confirm the non-detection of pulsation, we generated pulse 
profiles by assuming 50~s (corresponding to $\sim$20 mHz peak in the PDS) and earlier 
reported 67~s (from $Tenma$ observation) as the spin period of the source. We defined 
pulse fraction as the ratio between the difference in the maximum and minimum intensities 
to the sum of the maximum and minimum intensities in the pulse profile. We estimated 
pulse-fraction from each of the pulse profiles obtained by assuming 50~s and 67~s as 
spin period of the source and found to be $\sim$1\%. The negligible values of pulse 
fraction indicate the non-detection of X-ray pulsation in the source. On the other 
hand, the observed weak and broad feature at $\sim$20~mHz confirmed the detection of
a quasi-periodic oscillation (QPO) in the X-ray source. The significance of QPO feature 
was determined by fitting the PDS with a power-law continuum along with a Gaussian 
function at QPO frequency and found that the detection was more than 3$\sigma$ level.


\begin{figure}
\centering
\includegraphics[height=3.2in, width=2.3in, angle=-90]{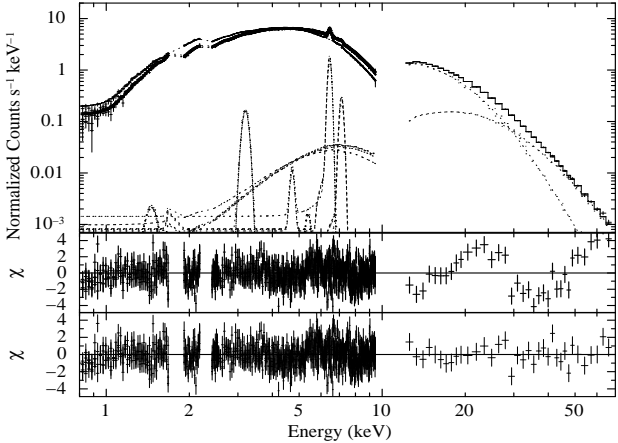}
\caption{Energy spectrum of 4U~1700-37 obtained with the XISs and PIN detectors 
of the $Suzaku$ observation, along with the best-fit model comprising  
a partial covering NPEX continuum model, three Gaussian functions for 
emission lines and a cyclotron absorption component. The middle and 
bottom panels show the contributions of the residuals to $\chi^{2}$ 
for each energy bin for the partial covering NPEX continuum model 
without and with cyclotron component in the model, respectively.}
\label{spec1}
\end{figure}

\begin{figure}
\centering
\includegraphics[height=3.2in, width=2.3in, angle=-90]{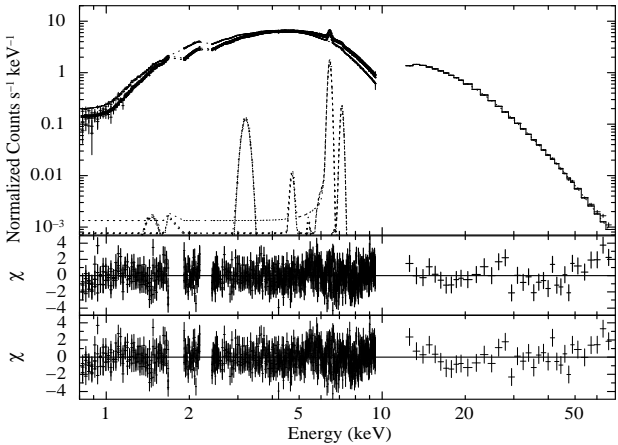}
\caption{Energy spectrum of 4U~1700-37 obtained with the XISs and PIN detectors 
of the $Suzaku$ observation, along with the best-fit model comprising  
a partial covering high energy cutoff power-law continuum model, three Gaussian 
functions for emission lines and a cyclotron absorption component. The middle and bottom 
panels show the contributions of the residuals to $\chi^{2}$ for each energy 
bin for the partial covering power law continuum model without and with cyclotron 
component in the model, respectively.}
\label{spec2}
\end{figure}

For spectral analysis, the source and background spectra, response matrices and effective 
area files for all instruments were generated by following the procedures described 
in previous section. After appropriate background subtraction, spectra from all
the detectors were fitted  simultaneously in 0.8-70 keV energy range using XSPEC v12.7 
package. Due to the presence of known Si and Au edge feature in XIS spectra, data 
in 1.7-1.9 keV and 2.2-2.4 keV energy ranges were ignored in the spectral fitting. 
The XIS spectra were binned by a factor of 6 from 0.8 to 10 keV whereas the 
PIN spectrum was binned by a factor of 2 up to 25 keV, a factor of 4 from 25 keV to 
50 keV and a factor of 6 from 50 to 70 keV. All the spectral parameters were tied 
together during the fitting except the relative normalization of detectors which 
were kept free. Standard continuum models for X-ray pulsars like high energy cutoff 
power-law (White, Swank \& Holt 1983), Fermi Dirac cutoff power-law (FDCUT; 
Tanaka 1986), NewHcut (a third order polynomial function with continuous 
derivatives; Burderi et al. 2000), cutoff power-law, NPEX (Makishima et al. 
1999), Thermal Comptonization model (CompTT; Titarchuk 1994) were applied in
the spectral fitting. However, the high energy cutoff power-law, NewHcut and 
NPEX model with partial covering component described the source source spectrum well. 

Addition of partial covering component to all three continuum models improved 
the spectral fitting yielding the reduced $\chi^2$ values from $>$4 to $<$2. In 
this model component, there are two different absorption components. One component
(equivalent hydrogen column density along the source direction) absorbs the entire 
spectrum, whereas the other component (inhomogeneously distributed matter close
to the X-ray source) absorbs the source spectrum partially. This model has been 
used to describe spectra of other HMXBs (Jaisawal et al. 2013; Pradhan et al. 2014) and 
Be/X-ray binary pulsars which show the presence of several absorption dips at various 
pulse phases during Type~I outbursts (Naik et al. 2011; Paul \& Naik 2011; Maitra 
et al. 2012; Naik et al. 2013 and references therein). Apart from these spectral
components in the continuum models, the iron fluorescence lines at 6.42 keV (Fe K$_\alpha$) 
and 7.1 keV (Fe K$_\beta$) were detected in the spectrum of 4U~1700-37. An emission 
line like feature at 3.19 keV was seen in the spectral residue of all three continuum 
models. Addition of a Gaussian component at $\sim$3.19 keV to above three continuum 
models improved the spectral fitting further. The line at $\sim$3.19 keV was identified 
as the fluorescence emission from S~XV as seen in EXO~2030+375 (Naik et al. 2013) or 
Ar K$_\beta$.


\begin{table*}
\centering
\caption{Best-fit parameters obtained from the spectral fitting of $Suzaku$ 
observation of 4U~1700-37 with 90\% errors. Model-1 : Partial covering NPEX 
model with Gaussian components, Model-2 : Partial covering NPEX model 
with Gaussian components and cyclotron absorption line, Model-3 : Partial 
covering high energy cutoff power-law model with Gaussian components, Model-4 : 
Partial covering high energy cutoff power-law model with Gaussian components 
and cyclotron absorption line, Model-5 : Partial covering NewHcut model with 
Gaussian components and Model-6 : Partial covering NewHcut model with Gaussian 
components and cyclotron line.}
\begin{tabular}{lllllll}
\hline
Parameter      		&\multicolumn{6}{|c|}{Value} 	 \\
                                       &Model-1             &Model-2            &Model-3           &Model-4      &Model-5           &Model-6  \\
\hline
N$_{H1}$$^a$    &1.9$\pm$0.1       &2.0$\pm$0.1         &2.2$\pm$0.3	    &2.2$\pm$0.1    &2.2$\pm$0.1  &2.2$\pm$0.1 \\
N$_{H2}$$^b$   &4.1$\pm$0.1       &4.5$\pm$0.2      &4.9$\pm$0.1	    &4.9$\pm$0.1    &4.5$\pm$0.1  &4.8$\pm$0.2\\
Cov. Fraction                      &0.6$\pm$0.1       &0.6$\pm$0.1      &0.7$\pm$0.1	    &0.7$\pm$0.1	  &0.7$\pm$0.1  &0.7$\pm$0.1 \\
Photon index                         &0.2$\pm$0.1       &0.3$\pm$0.1      &1.0$\pm$0.1	    &1.0$\pm$0.1    &0.9$\pm$0.1  &0.9$\pm$0.1\\
E$_{cut}$ (keV)	             &7.5$\pm$0.1       &8.8$\pm$1.5      &7.1$\pm$0.1      &7.1$\pm$0.1    &7.0$\pm$0.2  &7.0$\pm$0.2\\
E$_{fold}$  (keV)                  &--                  &  --               &19.1$\pm$0.1     &19.7$\pm$0.4   &18.6$\pm$0.4 &19.0$\pm$0.5\\
Fe K$_\alpha$ line\\
Line energy (keV)             &6.46$\pm$0.01     &6.46$\pm$0.01    &6.46$\pm$0.01    &6.46$\pm$0.01 &6.46$\pm$0.01 &6.46$\pm$0.01\\
Eq. width  (eV)        &81$\pm$2		        &82$\pm$3           &75$\pm$2           &75$\pm$2         &77$\pm$2       &77$\pm$2 \\

Fe K$_\beta$ line\\
Line energy (keV)              &7.13$\pm$0.01     &7.13$\pm$0.01    &7.15$\pm$0.01    &7.15$\pm$0.01  &7.14$\pm$0.01 &7.14$\pm$0.01\\
Eq. width  (eV)         &21$\pm$2		        &22$\pm$2           &14$\pm$1           &14$\pm$1         &20$\pm$2       &20$\pm$2\\
Cyclotron line\\
Line energy (keV)                   &--                  &38.9$\pm$3.2     &--                 &38.9$^*$     &--             &38.9$^*$ \\
Width (keV)          &--             &19.3$^{+6.1}_{-4.3}$     &-- 	          &9.8$^{+8.1}_{-5.1}$  &--      &7.3$^{+8.3}_{-5.5}$\\
Depth 	               &--                  &0.4$\pm$0.1      &--                 &0.1$\pm$0.1           &--      &0.1$\pm$0.1  \\
Flux$^c$ (1-10 keV)                    &2.1$\pm$0.1       &2.1$\pm$0.1      &2.1$\pm$0.1      &2.1$\pm$0.1    &2.1$\pm$0.1  &2.1$\pm$0.1\\
Flux$^c$ (10-70 keV)                   &5.6$\pm$0.3       &5.7$\pm$0.7      &5.6$\pm$0.2      &5.6$\pm$0.1    &5.6$\pm$0.2  &5.6$\pm$0.2\\ 
Norm. Const.$^d$            &1/1/1.04/0.98/1.02  &$----$  &1/1/1.04/0.98/1.02   &$----$  &1/1/1.04/0.98/1.01 &$----$ \\
 $\chi^2$ (dofs)                       &1551 (888)          &1363 (885)         &1389 (888)		      &1375 (886)       &1349 (887)      &1343 (885)\\
\hline
\end{tabular}
\\
\flushleft
$^a$ : Equivalent hydrogen column density in the source direction (in 10$^{22}$ atoms cm$^{-2}$ units),\\ 
$^b$ : Additional hydrogen column density (10$^{22}$ atoms cm$^{-2}$ units), \\$^c$ : in 10$^{-9}$  ergs 
cm$^{-2}$ s$^{-1}$ unit. \\
$^d$: Quoted relative normalization constants are for XIS-0, XIS-1, XIS-2, XIS-3 and PIN, respectively. 
The values remain same while adding CRSF with the respective continuum models (Model-1, Model-3 and Model-5).\\
$^*$ : The values were fixed at the value obtained from spectral fitting with Model-2.
\label{spec_par}
\end{table*}

In contrast to earlier findings, the thermal component (soft X-ray excess)  
was not seen in the spectrum during the {\it Suzaku} observation of 4U~1700-37. 
{\it XMM-Newton} observations of the source, however, showed the presence 
of soft X-ray excess over the continuum model in 0.22-0.31, 0.72-0.79 and 
0.07-0.17 orbital phase ranges. Apart from the eclipse phase when two soft 
X-ray excess components were detected, the source showed a single and weak 
soft X-ray excess during other orbital phases (van der Meer et al. 2005). 
Though the source was observed with {\it XMM-Newton} at four epochs, only 
one observation (in 0.48-0.59 phase range) overlaps partly with the binary 
orbital phase covered during {\it Suzaku} observation. The non-detection of 
soft X-ray excess during {\it Suzaku} observation of the source, therefore, 
can be explained as due to the relatively low sensitivity of {\it Suzaku} 
instruments compared to that of {\it XMM-Newton} to detect the weak soft 
component in the spectrum. 

Apart from the soft X-ray excess, earlier reported absorption like
feature at $\sim$37 keV in the source spectrum (Reynolds et al. 1999) was 
also marginally seen in the spectral residuals of all three continuum models 
in the present work. However, the absorption feature was clearly detected when 
the source spectrum was fitted with partial covering NPEX continuum model. 
Addition of cyclotron absorption component to the partial covering NPEX 
continuum model improved the spectral fitting yielding better value of
reduced $\chi^2$ which was decreased from 1.75 to 1.54. Energy and width 
of the cyclotron absorption feature were found to be $\sim$39 keV and 
$\sim$19 keV, respectively. The cyclotron absorption component was added 
to the partial covering high energy cutoff power-law and NewHcut models 
yielding almost identical results. The parameters obtained from the spectral 
fitting of {\it Suzaku} observation of 4U~1700-37 are given in Table~1. 
The values of relative instrument normalizations of four XISs and HXD/PIN 
are also given in the table and found to be comparable to that obtained during 
the detector calibrations. The energy spectra of the source are shown in 
Figures~\ref{spec1} and ~\ref{spec2} along with the best-fit models of partial 
covering NPEX model and partial covering high energy cutoff power-law model, 
respectively. The middle and bottom panels in each figure show the residuals 
to the best-fit models without and with the addition of cyclotron absorption 
line in the continuum model, respectively. 

To check the statistical significance of the absorption feature, F-test routine of $IDL$, 
$mpftest$\footnote{http://www.physics.wisc.edu/$\sim$craigm/idl/down/mpftest.pro}, 
was applied on the $\chi^2$. As in case of 4U~1909+07 (Jaisawal et al. 2013), 
probability of chance improvement (PCI) was evaluated by considering the $\chi^2$ 
without and with the cyclotron absorption component in continuum models (Press et 
al. 2007). The estimated PCI was found to be 3\%, 46\% and 49\% after adding the cyclotron 
component in the partial covering NPEX model, the partial covering high energy 
cutoff power-law  model and the partial covering NewHcut model, respectively. 
At such high PCI values (46\% and 49\%), the detection of cyclotron absorption 
feature is statistically insignificant in spectral fitting with partial covering 
high energy cutoff power-law  and partial covering NewHcut models. Though the PCI 
value for partial covering NPEX model (3\%) suggest the detection of cyclotron
absorption feature in the source, the broad width of the feature ($\sim$19 keV) 
makes the detection tentative. Considering the different values of PCI obtained 
for different models, the use of F-test in checking the statistical significance of 
the presence of cyclotron absorption component in the spectrum of 4U~1700-37 
(present case) may not be reliable enough. Observations with high sensitive hard 
X-ray detectors for long exposures can confirm the presence/absence of cyclotron 
resonance scattering feature in 4U~1700-37.

\begin{figure*}
 \centering
 \includegraphics[height=5.3in, width=3.7in, angle=-90]{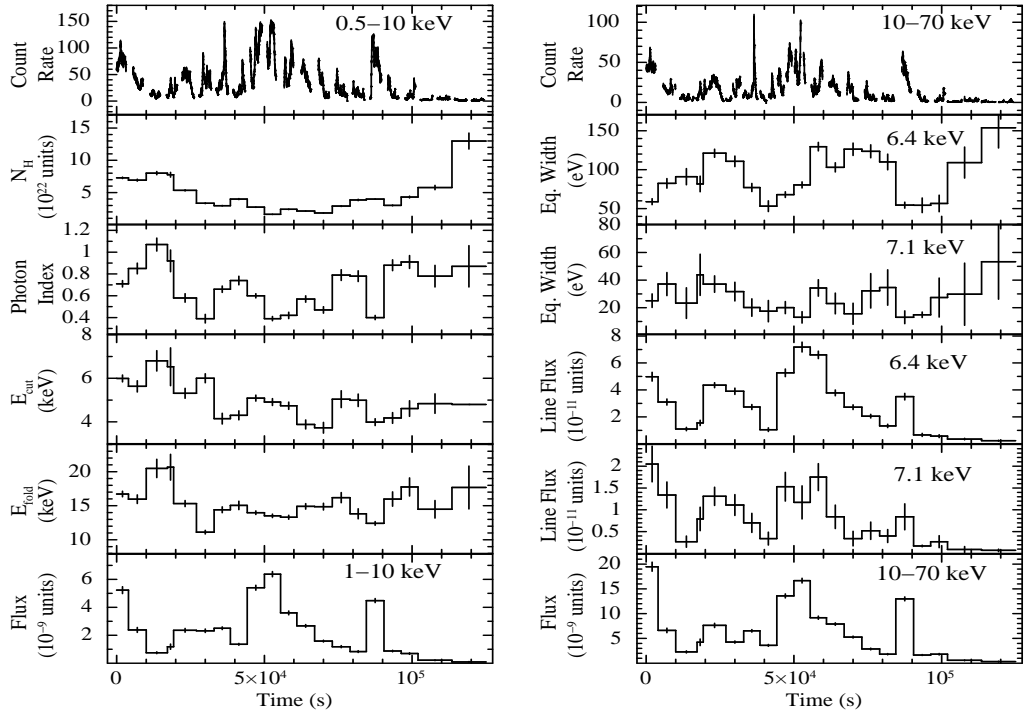}
 \caption{Spectral parameters obtained from the time-resolved spectroscopy 
for $Suzaku$ observation of 4U~1700-37. The top panels in both the sides 
show light curves of 4U~1700-37 in 0.5-10 keV (XIS-0) and 10-70 keV (HXD/PIN) 
energy ranges. The values of $N_H$, power-law photon index, cutoff ($E_{cut}$) 
and folding energy ($E_{fold}$) are shown in second, third, fourth and fifth 
panels in left side, respectively. The iron emission line parameters such as
the equivalent widths and flux in for 6.4 keV and 7.1 keV iron emission lines
are shown in second, third, fourth and fifth panels in right side, respectively. 
The source flux in 1-10 keV (left side) and 10-70 keV (right side) are shown
in bottom panels.The errors shown in the figure are estimated for 90\% 
confidence level.}
\label{sp}
 \end{figure*}

 \subsection{Time-resolved spectroscopy}
   
During the $Suzaku$ observation of 4U~1700-37, observed source flux was found 
to be highly variable at different time scales. Several flare like episodes 
lasting for $\sim$10 ks and low flux segments were seen in soft and hard X-ray 
light curves (Figure~\ref{lc}). To probe the changes in spectral parameters 
during these flaring episodes and low flux segments at such short intervals, 
we divided the entire observation into 20 segments as marked in the top panel 
of Figure~\ref{lc}. As mentioned earlier, source spectra for these 20 segments 
were extracted from all four XISs and PIN detectors. For time-resolved 
spectroscopy, we used same background spectra and response matrices for 
corresponding detectors that were used for time-averaged spectroscopy. As 
all three models used in time-averaged spectroscopy well fitted the source 
spectrum, we choose one of the model e.g. the high energy cutoff power-law 
model to fit the time-resolved spectra. Iron K$_\alpha$ and K$_\beta$ 
lines were detected in each of the 20 time-resolved spectra. The best-fit
spectral parameters (with 90\% errors) obtained from the simultaneous spectral
fitting of each of the segments are plotted in Figure~\ref{sp} along with XIS-0
and PIN light curves in left and right top panels, respectively.

The equivalent hydrogen column density N$_H$ was found to be high 
(7$\times$10$^{22}$~cm$^{-2}$) in the beginning of observation. However, 
the values of N$_H$ gradually decreased to a low value beyond which again 
showed gradual increase during the observation. The systematic and smooth 
variation of N$_H$ irrespective of source intensity during the {\it Suzaku} 
observation suggested the orbital dependence of the matter distribution in 
4U~1700-37. This can be confirmed with further long observations of the 
source with upcoming observatories such as ASTROSAT. A sharp increase in 
the value of N$_H$ (13$\times$10$^{22}$~cm$^{-2}$) was observed during the 
extended low flux segment towards the end of the observation (after $\sim$0.63 
orbital phase). The power-law photon index was found to be variable e.g. higher 
during the low flux segments compared to the flaring episodes. 

\begin{figure*}
\vskip 8.8cm
\includegraphics{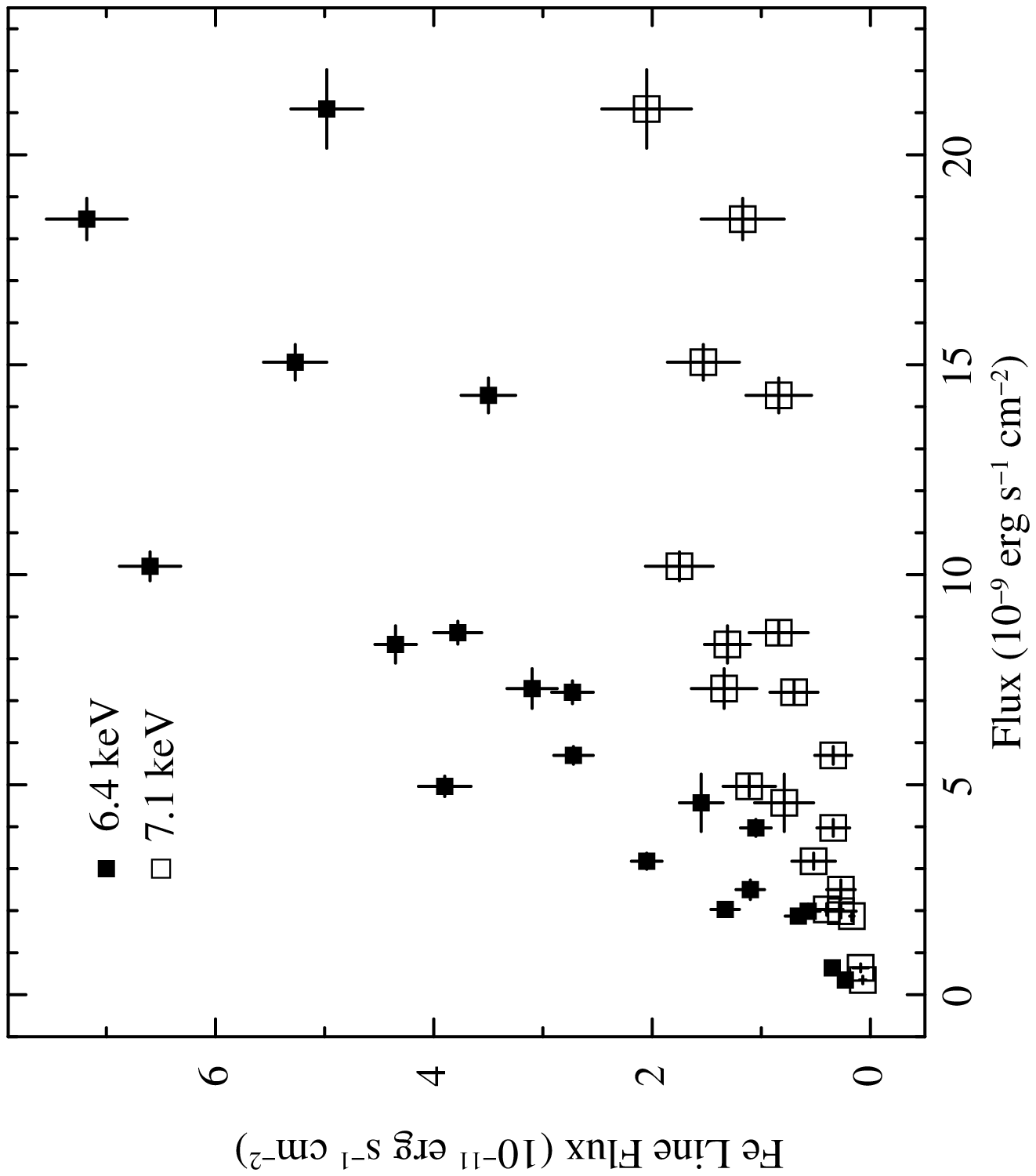}
\includegraphics{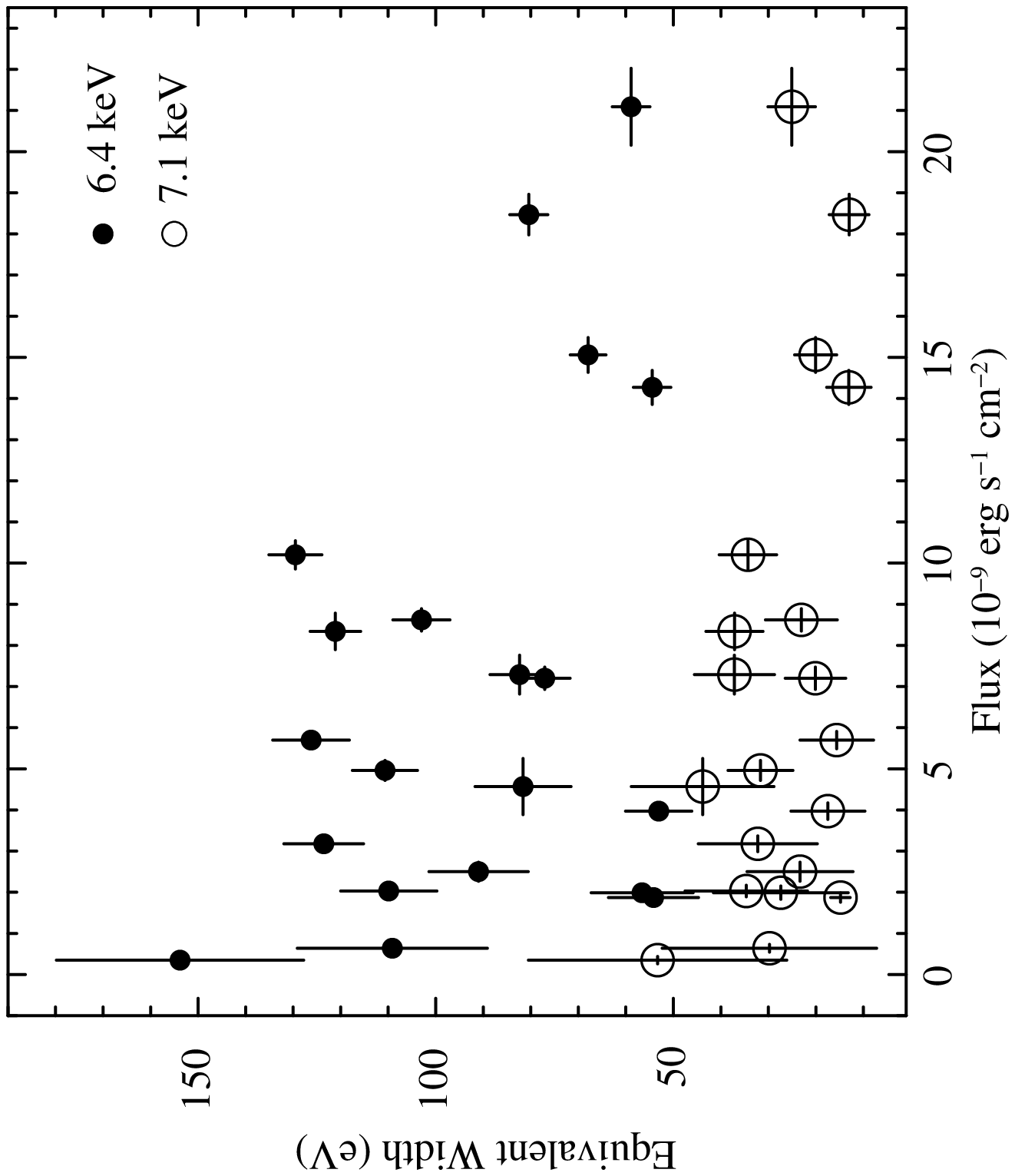}
\caption{Change in the 6.4 keV and 7.1 keV iron emission line flux (left panel) 
and equivalent widths (right panel) with respect to the estimated flux in 8-70 keV
energy range.}
\label{Fe-line}
\end{figure*}

\begin{figure}
\centering
\includegraphics[height=3.1in, width=2.6in, angle=-90]{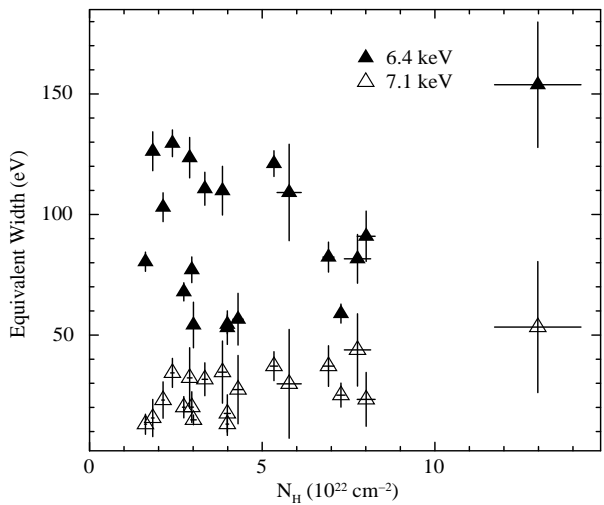}
\caption{Change in the 6.4 keV and 7.1 keV iron line equivalent widths with 
respect to the column density during the {\it Suzaku} observation.}
\label{eqw-nh}
\end{figure}

Flux of Iron K$_\alpha$ and K$_\beta$ emission lines were found to vary with 
the source flux whereas the corresponding equivalent widths showed the opposite 
trend. Source flux as well as the flux of iron emission lines were found to be 
low during the extended low flux segment at the end of the observation. The 
variation in the iron line parameters (flux and equivalent width) with the 
absorbed source flux in 8-70 keV are plotted in Figure~\ref{Fe-line}. Though
the flux of both the emission lines increased along with the source flux, 
flux of K$_\alpha$ line was found to increase faster compared to that of the
K$_\beta$ line. However, the equivalent width of both the lines showed no
systematic variation with the source flux though the values were higher at low 
flux level. Dependence of emission line flux and equivalent width with the source 
flux indicates the fluorescence origin of the lines from the matter near by the 
neutron star. Variation of iron emission line flux and equivalent width with
hard X-ray continuum flux (8-70 keV) in 4U~1700-37 are found to be similar to
that found in LMX~X-4 and Her~X-1 (Naik \& Paul 2003). In case of LMC~X-4 and
Her~X-1, the change in iron line parameters with continuum flux was interpreted
as due to the presence of precessing tilted accretion disk causing modification
in the geometry and visibility of iron line emitting region in the binary systems.
To investigate the geometry in 4U~1700-37 binary system, we plotted the equivalent 
widths of 6.4 keV and 7.1 keV iron lines with the observed column density 
in Figure~\ref{eqw-nh}. The equivalent widths of both the lines are found to be
marginally variable with the column density. Similar kind of variation of equivalent 
width (below of 200~eV) with absorption column density (order of 10$^{22}$~cm$^{-2}$) 
was seen in Her~X-1 and Vela~X-1 (Figure~8 of Makishima et al. 1986). In such 
configuration, the X-ray source is expected to be surround by inhomogeneously distributed
absorbing material that covers a fraction of radiation along the line of sight.

\section{Discussion}

The continuum X-ray emission in the neutron star X-ray binaries is understood 
to be emitted through the inverse Comptonization of the photons originated 
from the magnetic poles as well as within the accretion column of the neutron star 
(Becker \& Wolff 2007). The emitted photons interact with the materials present 
in the surroundings of the neutron star. The effect of interactions can be seen in 
the observed spectrum in the form of photo-electric absorption, soft excess emission, 
line emissions and cyclotron absorption features. Detection of soft excess and 
line emissions can reflect the reprocessing of hard X-ray photons with matter 
(neutral or ionized plasma) in accretion disk, accretion column etc. The effect 
of the neutron star magnetic field can be seen in the X-ray spectrum through the 
interaction of quantized electrons with the source photons. Depending on the strength
of the neutron star magnetic field, the electrons are quantized in to Landau levels 
through the relation E$_a=11.6 B_{12} (1+z_g)^{-1}$ (keV), where $z_g$ is gravitational 
red-shift and B$_{12}$ is magnetic field strength in units of 10$^{12}$~Gauss. The 
absorption like feature, cyclotron resonance scattering feature (CRSF), is detected 
generally in 10-100 keV spectrum. In the present case of high mass X-ray binary 
4U~1700-37, we detected an absorption like feature at $\sim$39~keV, as was earlier 
reported in $BeppoSAX$ spectrum of the source (Reynolds et al. 1999). During 
the $BeppoSAX$ observation, the cyclotron absorption feature was detected when the
source spectrum was fitted with a high energy cutoff power-law model. However, in 
the present study, the cyclotron absorption feature was detected when the spectrum
was fitted with a NPEX model. Though the detections of CRSF during {\it Beppo}SAX
observation as well as {\it Suzaku} observation did not provide any concluding
results, the presence of weak absorption-like feature at $\sim$39 keV in both the
cases can not be entirely ruled out. Assuming the detected feature as CRSF in 
4U~1700-37, the magnetic field of the X-ray source (neutron star) in the binary 
system was estimated to be $\sim$3.4$\times$10$^{12}$~Gauss. The presence of 
the cyclotron feature can be confirmed by using data from future observations with 
long exposure and good hard X-ray spectroscopic instruments such as $NuSTAR$, 
$ASTROSAT$ \& $Astro-H$ missions.

4U~1700-37 was found to be significantly variable during the $Suzaku$ observation. 
Such kind of flux variability are seen in other wind-accreting high mass X-ray binaries 
such as Vela~X-1 (Kreykenbohm et al. 2008), Cen~X-3 (Naik et al. 2011), 4U~1907+09 
(River et al. 2010) etc. The source luminosity in these wind-fed systems depends on 
the density and velocity of stellar wind as  $L_x\propto\rho v^{-3}$ (Bondi \& Hoyle 1944).  
Any fluctuations either in density or velocity can produce the variation in luminosity. 
Flux variability in time scales of kilo-seconds as seen in 4U~1700-37, were seen in 
Vela~X-1 and was explained on the basis of clumpy wind with the fluctuating density
causing variation in the accretion rate (Kreykenbohm et al. 2008; Odaka et al. 2013). 
Presence of low and high flux levels were also seen in the pulsar 4U~1907+09 during
$Suzaku$ observations in 2006 and 2007 (Rivers et al. 2010). Though the low flux 
levels were present in the light curves of both the observations, the one present 
in 2006 observation was consistent with earlier observations of 4U~1907+09 and 
interpreted as due to change in whole continuum rather than obscuration/absorption 
of X-rays due to the presence of additional matter as in later case (Rivers et al. 
2010). The argument of presence of clump of matter causing low flux levels in the 
light curves during 2007 observation was supported by the enhancement in the value 
of absorption column density along the line of sight. However, in the present 
study, the variation of N$_H$ was marginal during flare and non-flare durations. 
Significantly high value of N$_H$ after orbital phase $\sim$0.6 compared to rest
of the segments of the {\it Suzaku} observation confirmed the finding from {\it
EXOSAT} observation of the source. This high N$_H$ value segment (beyond orbital
phase of $\sim$0.6) was interpreted as due to the passage of the accretion 
wake between the neutron star and the observer. As the neutron star moves away, 
the accretion wake which trails behind the neutron star during the whole 
orbit, crosses the line of sight of the observer at $\sim$0.6 orbital phase 
yielding significantly high value of absorption column density.

Taam \& Fryxell (1989) performed the simulation to understand the interaction 
between asymmetric accretion flow from OB stars onto the neutron stars. The results 
showed that a temporary disk can be formed during the interaction of accretion flow 
with the shock in accretion wake region. The destruction of temporary accretion disk 
is associated with the reversal of storage accretion flow that increases the mass 
accretion rate. The flow reversal occurs in the range of few hours that generates 
the flares of 15 m to 1 hr time scales as seen in several segments (such as 3, 4, 7, 
10 etc.) of present {Suzaku} observation. The ``flip-flop instability'' in the 
accretion disk can possibly also explain short time flaring activities as observed 
in 4U~1700-37 (Matsuda et al. 1991). However, in an alternate scenario, the 
hydrodynamics simulation results for wind-fed sources showed the formation of 
non-steady accretion wake consisting the dense filaments of compressed gas where 
the density reaches $\sim$100 times more compared to undisturbed stellar wind 
(Blondin et al. 1990). Accretion of these filaments with fluctuating density may 
generate the abrupt variation in the X-ray luminosity as observed in the present 
case.

$Suzaku$ observation of 4U~1700-37 was taken during out of eclipse of binary.
However, an eclipse-like low flux segment was observed towards the end of the
observation in 0.63-0.73 orbital phase range. During this segment, an increase 
in column density was found. The source flux and line flux of both the iron 
emission lines were decreased to minimum values compared to the rest of the 
observation. The presence of dense matter in this orbital phase range can be 
the possible reason for the eclipse like segments. Such type of eclipse-like 
segments (quiescence period) was also observed during $Chandra$ observation of
4U~1700-37 around $\sim$0.68 orbital phase (Boroson et al. 2003). A significant 
increase in the column density after phase 0.5 was also reported earlier during 
the $Copernicus$ observation of 4U~1700-37 (Mason et al. 1976). During $EXOSAT$ 
observation, an increasing column density was also noticed after 0.6 orbital phase
of the binary (Haberl et al. 1989). The sharp increase in the column density at 
above orbital phase range can be interpreted as the formation of accretion wake 
as observed in Vela~X-1 (Blondin et al. 1990). Haberl et al. (1994) also reported 
the presence of accretion wake based on the temperature difference observed in the 
soft excess component. However, the spectroscopic evidences confirmed the formation 
of accretion wake at late orbital phases of binary that blocks the continuum and 
produces the eclipse like segments (Kaper et al. 1994) which is seen in the present 
case of 4U~1700-37.

In summary,  the observed flux variability at ks time scale during $Suzaku$ 
observation of 4U~1700-37 can be explained on clumpiness in the stellar wind 
which may cause fluctuation in mass accretion rate. However, the 
instability in temporary disk or flip flop instability in accretion disk can 
also produce the flares on short time intervals as seen in light curves between 
0.29-0.63 orbital phase. The extended low flux eclipse-like segment observed 
towards the end of the observation is interpreted as due to the presence of 
accretion wake.

\section*{Acknowledgments}
The authors would like to thank the referee for his/her constructive comments 
and suggestions that improved the contents of the paper. The research work at 
Physical Research Laboratory is funded by the Department of Space, Government 
of India. The authors would like to thank all the members of the Suzaku for 
their contributions in the instrument preparation, spacecraft operation, software 
development, and in-orbit instrumental calibration. This research has made use of 
data obtained through HEASARC Online Service, provided by the NASA/GSFC, in support 
of NASA High Energy Astrophysics Programs.

\end{document}